\documentclass[twocolumn,amsmath,amssymb,floatfix]{revtex4}

\usepackage{graphicx}
\usepackage{dcolumn}
\usepackage{bm}

\graphicspath{{Figures/}}
\setkeys{Gin}{width=\linewidth}

\begin{document}

\title{Single-hole transistor in p-type GaAs/AlGaAs heterostructures}

\author{Boris Grbi\'{c}$^{*}$, Renaud Leturcq$^{*}$,  Klaus
  Ensslin$^{*}$, Dirk Reuter$^{+}$, and Andreas D. Wieck$^{+}$}

\affiliation{$^{*}$Solid State Physics Laboratory, ETH Zurich,
  8093 Zurich, Switzerland,\\$^{+}$Angewandte Festk\"{o}rperphysik,
Ruhr-Universit\"{a}t Bochum, 44780 Bochum, Germany}

\begin{abstract}
A single-hole transistor is patterned in a p-type, C-doped
GaAs/AlGaAs heterostructure by AFM oxidation lithography. Clear
Coulomb blockade resonances have been observed at T$_{hole}$=300
mK. A charging energy of $\sim$ 1.5 meV is extracted from Coulomb
diamond measurements, in agreement with the lithographic
dimensions of the dot. The absence of excited states in Coulomb
diamond measurements, as well as the temperature dependence of
Coulomb peak heights indicate that the dot is in the multi-level
transport regime. Fluctuations in peak spacings larger than the
estimated mean single-particle level spacing are observed.

\end{abstract}

\maketitle


Quantum dots implemented in GaAs heterostructures represent
promising candidates for the experimental realization of quantum
computation \cite {Loss98}, as well as spintronics devices \cite
{Wolf01}. However, research based on electronic transport through
such small conducting islands was, so far, exclusively focused on
quantum dots defined on n-type GaAs heterostructures (for a
review, see \cite {Kouwenhoven97}). In this paper we report about
Coulomb blockade (CB) measurements in a single-hole transistor,
defined on a p-type carbon doped GaAs heterostructure.

The interest in low dimensional hole-doped systems arises
primarily from the fact that spin-orbit as well as carrier-carrier
Coulomb interaction ($E_{int}$) effects are more pronounced in
such systems compared to the more established electron-doped
systems. The main reason for this is that holes have much higher
effective masses than electrons, and thus a smaller Fermi energy.
This should allow the investigation of novel regimes with much
higher interaction parameter $r_s=E_{int}/E_F$.

However, stronger spin-orbit interaction in hole-doped systems
significantly reduces spin relaxation times in bulk p-doped GaAs
systems. It was also shown \cite {Schneider04} that spin
relaxation of holes confined into quantum wells is much slower
than in the bulk case, but still several orders of magnitude
faster than electron spin relaxation. This was one of the main
reasons why hole systems received little attention in efforts to
utilize spin in quantum information technologies. Recently, Bulaev
and Loss predicted \cite {Bulaev05}, that further confinement of
holes into quantum dots can significantly increase the relaxation
time $T_1$ of hole spins, so that it can be comparable, or even
larger than that of electron spins.

Another reason why low-dimensional hole systems have received less
attention is due to technological difficulties to fabricate stable
p-type structures. Nanodevices fabricated on p-type GaAs with
conventional split-gate techniques show significant gate
instabilities \cite {Zailer94}. This is presumably due to the fact
that metallic Schottky barriers on p-type GaAs are more leaky than
on n-type GaAs \cite {Williams}. Our experience with metallic
split-gates on p-type GaAs also shows that they display hysteretic
behavior, making them unsuitable for high-precision tuning of the
devices. Therefore, we decided to employ another technique,
namely, Atomic Force Microscope (AFM) oxidation lithography \cite
{Held98, Rokhinson02} to define nanostructures on two-dimensional
hole gases (2DHG).

The crucial point for the implementation of AFM oxidation
lithography technique is that the 2DHG is located close to the
sample surface. In case of our heterostructure, the 2DHG resides
at the interface 45 nm below the sample surface. The
heterostructure itself consists of 5 nm undoped GaAs cap layer,
followed by a 15 nm thick, homogeneously C-doped layer of AlGaAs,
which is separated from the 2DHG by a 25 nm thick, undoped AlGaAs
spacer layer. Before performing AFM lithography, the sample was
characterized by standard magnetotransport measurements at 4.2 K
and the following values are obtained for its density and
mobility: n = 4$\times$10$^{11}$ cm$^{-2}$, $\mu$=120'000
cm$^2$/Vs . The effective masses of the two spin-split subbands in
C-doped GaAs heterostructures are $m_1=0.34 $ $m_e$ and $m_2$ =
0.53 $m_e$ \cite{Grbic04}. The typical values for the interaction
parameter in our system are determined to be $r_s>5$.

\begin{figure}[h]
  \begin{center}
    \includegraphics[width=0.96\linewidth]{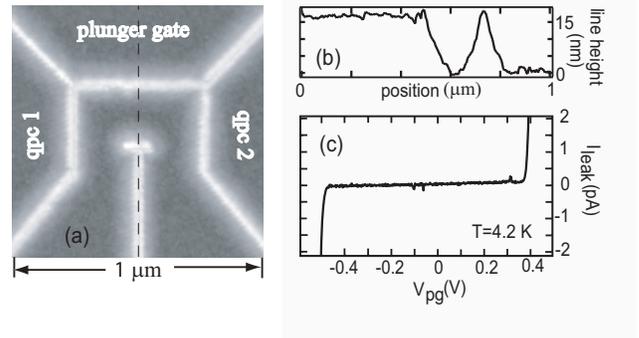}
    \caption{a) AFM micrograph of the quantum dot with designations of the gates: qpc 1 and qpc 2 are the gates for tunning the coupling of the dot to the source and drain, while the plunger gate serves to tune the number of holes in the dot. Bright oxide lines fabricated by AFM oxidation lithography lead to insulating barriers in the 2DHG. (b) Height profile along the dashed line in Fig. 1a - the oxide segments are 15-18 nm high. (c) Test of the insulating behavior of the oxide line at T=4.2 K : A voltage is applied to the plunger gate and the total current to all other gates, which are kept grounded, is measured. For this sample the oxide lines are insulating for the applied voltages in the range [-500 mV, +400 mV].}
    \label{fig1}
  \end{center}
\end{figure}

It is important to mention that C acts as an acceptor on the (100)
plane \cite{Wieck00}, and thus the anisotropy in the hole gas
formed in this plane is significantly suppressed compared to the
case of Si doped (311) heterostructures \cite {Grbic_unpubl}. Thus
the functionality of devices patterned on C-doped GaAs wafers as
well as the interpretation of the transport experiments in these
devices should not be dependent on the particular orientation of
the device with respect to the wafer.

An AFM micrograph of the quantum dot is shown in Fig. 1(a). The
lithographic dimensions of the dot are $430\times170$ nm$^2$,
while the width of both quantum point contacts (qpc) is $\sim 140$
nm. The height profile in Fig. 1(b) along the dashed line in (a)
indicates excellent topological homogeneity and quite constant
height of the oxide lines.

We demonstrate that for a 2DHG 45 nm below the sample surface AFM
written oxide lines with a height of 15-18 nm completely deplete
the 2DHG beneath at 4.2 K. Voltages in the range [-500 mV, +400
mV] can be applied between separated regions without any
significant leakage current across the oxide line (Fig. 1(c)).
This allows us to make in-plane gates for defining nanostructures
with satisfying tunability.

The transport measurements in the dot have been performed in a
$^3$He/$^4$He dilution refrigerator at a base temperature of
$\sim$ 50 mK. We have measured the two-terminal conductance
through the dot by applying either a small dc or ac bias voltage
V$_{sd}$ between source and drain, and measuring the current
through the dot with a resolution better than 50 fA.

\begin{figure}
  \begin{center}
    \includegraphics{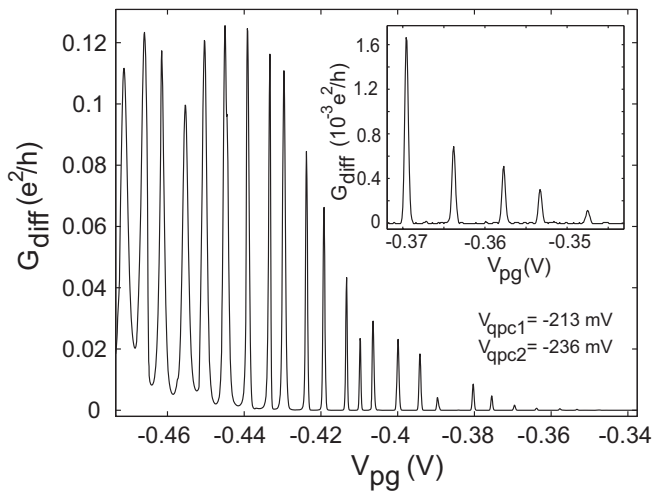}
    \caption{Differential conductance through the dot as a function of plunger gate voltage $-$ clear Coulomb resonances are observed. Measurements are performed in the dot configuration:  V$_{qpc1}$ = -213 mV, V$_{qpc2}$ = -236 mV, with symmetrically  applied AC source-drain bias of 20 $\mu$V and frequency of 31 Hz at the base temperature of 50 mK. Inset: Blow up of the weak coupling regime.
}
    \label{fig2}
  \end{center}
\end{figure}

The gates qpc1 and qpc2 are tuned in the configuration V$_{qpc1}$
= -213 mV, V$_{qpc2}$ = -236 mV in which the dot is symmetrically
coupled to source and drain. The differential conductance of the
dot is measured as a function of the plunger gate voltage.
Pronounced Coulomb resonances are observed (Fig. 2). It is
important to note that the dot closes when the value of the
plunger-gate voltage increases $-$ this is a clear indication that
we measure hole transport. In this configuration of the gate
voltages the peak positions were stable in twenty consecutive
plunger-gate sweeps within an accuracy of 0.1 mV. However, it is
important to mention that not every gate configuration shows such
stability - in certain configurations charge rearrangements make
reproducible measurements difficult.

We now focus on the weak coupling regime shown in the inset of
Fig. 2. Each of these five resonances is fitted both with an
expression for a thermally broadened Coulomb blockade peak in the
multi-level transport regime and a coupling broadened Lorentzian
peak \cite{Beenakker91} (Fig. 3). In all cases the thermally
broadened resonance fits significantly better to the data than a
coupling broadened resonance, indicating that the dot is really in
the weak coupling regime and that the peak broadening is
determined by temperature rather than coupling. As a fit parameter
we obtain FWHMs for each of these five peaks and they are in the
range from 440 $\mu$V to 480 $\mu$V, which correspond to hole
temperatures in the range T$_{hole}=300-330$ mK.

\begin{figure}
  \begin{center}
    \includegraphics[width=0.8\linewidth]{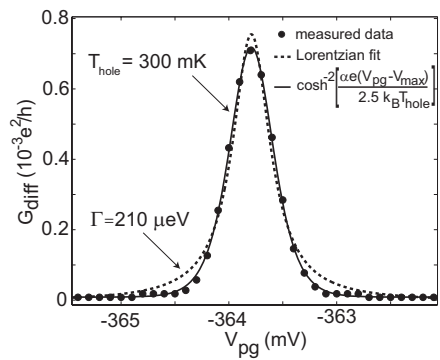}
    \caption{Peak at V$_{pg}$ = -363.8 mV  (circles) fitted to a  thermally broadened Coulomb blockade resonance (full line) and to a Lorentzian (dashed
    line). In the thermally broadened fit $\alpha$ is the lever-arm
    of the plunger gate, while V$_{max}$ and T$_{hole}$ are fitting
    parameters.
}
    \label{fig3}
  \end{center}
\end{figure}

Coulomb diamond measurements, i.e., measurements of the
differential conductance as a function of bias voltage V$_{bias}$
and plunger gate voltage V$_{pg}$, are performed in the weak
coupling regime, and the results are shown in Fig. 4. The uniform
size of the diamonds indicates that all confined holes reside in
one single potential minimum rather than occupying several
disconnected or tunnel-coupled potential minima, as it was
reported for a p-type SiGe quantum dot \cite{Dotsch01}. From the
extent of the diamonds in bias direction we estimate a charging
energy of the dot to be $E_{C}\approx1.5$ meV, while the lever-arm
of the plunger gate is $\alpha\approx0.26$. This charging energy
corresponds to a capacitance of the dot
$C=e^2/E_{C}\approx1.1\times10^{-16}$ F. If we assume a disk-like
shape of the dot, the capacitance is given by
$C=8\varepsilon_0\varepsilon_rr$, where $r$ is the radius of the
dot. This allows us to estimate the electronic diameter of the dot
to be $\sim 230$ nm, which is in very good agreement with the
lithographic dimensions of the dot and indicates that the dot is
really formed in the region encircled by the oxide lines.

\begin{figure}
  \begin{center}
    \includegraphics{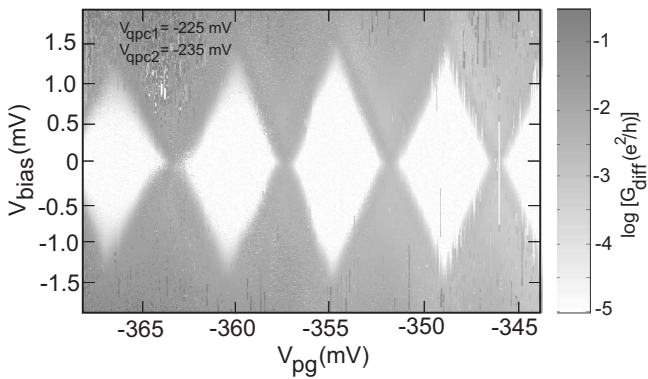}
    \caption{Coulomb diamonds in differential conductance, represented in a logarithmic gray scale plot (white regions represent low conductance). A DC bias is applied symmetrically across the dot and the current through the dot is measured. Differential conductance is calculated by numerical derivation. Charging energy of the dot is estimated to be $\sim$ 1.5 meV from this measurement. Measurements are performed in the dot configuration:  V$_{qpc1}$ = -225 mV, V$_{qpc2}$ = -235 mV.}
    \label{fig4}
  \end{center}
\end{figure}

It can be seen in Fig. 4 that in certain dot configurations,
low-frequency switching noise due to charge rearrangements in the
sample becomes quite expressed. Although for low biases applied
across the dot, the gate configurations can be found in which the
switching noise in current through the dot can be completely
suppressed, for large source-drain biases the switching noise
becomes more pronounced and is present in almost all dot
configurations.

In the case of the dot with steep potential walls, the mean
single-particle level spacing can be calculated as
$\Delta=2\pi\hbar^2/gm^*A$, where $g$ is the degeneracy of hole
states and $A$ is the electronic area of the dot. Due to the large
effective mass of holes ($m_1=0.34 $ $m_e$ and $m_2$ = 0.53
$m_e$), the mean single-particle level spacing in the dot is
estimated to be $\triangle\leq 15$ $\mu$eV, which is one order of
magnitude smaller than typical values in electron quantum dots.
Since we have $k_BT_{hole}\approx 25$ $\mu$eV for the estimated
hole temperature, the dot is in the regime where $\Delta \leq
k_BT_{hole}$. This explains why excited states cannot be resolved
in Coulomb diamond measurements.

We further explored the temperature dependence of the Coulomb
peaks and found that the peak amplitude does not decrease, but
rather increases as the temperature increases (Fig.5). This is
another indication that the dot is not in the single-level
transport regime, but in the intermediate multi-level transport
regime where several (about ten) single-particle levels
participate in transport  \cite {Kouwenhoven97}.

\begin{figure}
  \begin{center}
    \includegraphics[width=0.99\linewidth]{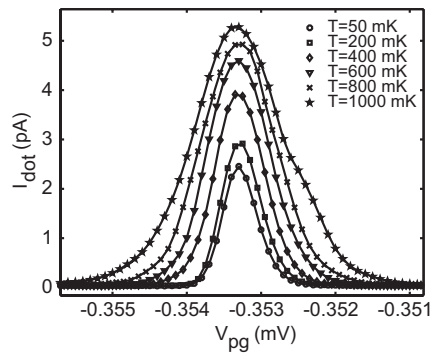}
    \caption{Temperature evolution of the Coulomb peak at V$_{pg}$ = -353.3 mV. Indicated are the bath
    temperatures.
}
    \label{fig5}
  \end{center}
\end{figure}

Despite the fact that the dot is in the multi-level transport
regime, fluctuations in peak spacings are observed. The positions
(in V$_{pg}$) of the five peaks shown in the inset of Fig. 2 are
respectively: -369.6 mV, -363.8 mV, -357.7 mV, -353.3 mV, and
-347.5 mV. Thus, the peak separations between consecutive peaks
are: 5.8 mV, 6.1 mV, 4.4 mV and 5.8 mV. These fluctuations in peak
separation, significantly larger than the estimated mean
single-particle level spacing in the dot, may indicate some
significance of stronger carrier-carrier interactions in hole
quantum dots \cite{Simmel99}.

Finally, we performed a second cool-down of the same dot in
another refrigerator where the thermal coupling of the sample to
the mixing chamber is better. Narrower Coulomb resonances were
obtained and the hole temperature extracted from the fitting of
the peaks in this case was in the range T$_{hole}=100-130$ mK.
This indicates, that the peak width is really determined by
temperature, and that there is no intrinsic mechanism in the
sample which widens the Coulomb peaks up to this energy range
\cite{Dotsch01}. Eventhough the hole temperature was lower in the
second than in the first cool-down, it was still not possible to
resolve excited states in Coulomb diamond measurements.

In conclusion, we fabricated a quantum dot on a p-type GaAs/AlGaAs
heterostructure by AFM oxidation lithography. Its functionality is
demonstrated by observing clear and reproducible Coulomb
resonances. From the Coulomb diamond measurements it was estimated
that the charging energy of the dot is $\sim1.5$ meV, which is
compatible with the lithographic dimensions of the dot. However,
due to the small single-particle level spacing in case of holes it
was not possible to resolve excited states in the dot. In order to
be able to investigate the single-particle level spectrum in hole
quantum dots, one has to significantly reduce both, the lateral
dimensions of the dot as well as the hole temperature. Conductance
peak spacings statistics are greatly affected by interaction
effects. Exploring these statistics in hole quantum dots in
single-particle level regime would bring new information about the
importance of carrier-carrier and spin-orbit interactions in low
dimensional systems.

It's our pleasure to acknowledge valuable discussions with R.
Schleser and D. Graf concerning the AFM oxidation lithography
technique and with T. Ihn and A. Gildemeister concerning the
measurement setup. Financial support from the Swiss National
Science Foundation, the German Science Foundation (DFG, GRK 384)
and the German Ministry for Science and Education (BMBF, 01BM451)
is gratefully acknowledged.


\begin{thebibliography}{99}
\bibitem{Grbic_unpubl}
We have tested several different C-doped (100) GaAs
heterostructures in order to find the one suitable for AFM
lithography. In all of these wafers the mobility anisotropy is
less than 25\%. This is significant improvement compared to the
more established Si doped (311) wafers where the mobility along
one of the main in-plain crystalographic directions is typically
2-3 times larger than in the other direction.
\end{thebibliography}
\end{document}